\input{aipcheck}

\documentclass[
    ,final            
  ]
  {aipproc}

\layoutstyle{6x9}

\begin{document}

\title{AGN FEEDBACK IN THE COMPACT GROUP OF GALAXIES HCG 62
-- as revealed by {\it Chandra}, {\it XMM} and {\it GMRT} data}


\author{Gitti M.}{
  address={Harvard-Smithsonian Center for Astrophysics,
60 Garden Street, Cambridge, MA 02138, USA}
  ,altaddress={Dept. of Astronomy, University of Bologna \& INAF OABo,
via Ranzani 1, I-40127 Bologna, Italy}
}

\author{O'Sullivan E.}{
  address={Harvard-Smithsonian Center for Astrophysics,
60 Garden Street, Cambridge, MA 02138, USA}
}

\author{Giacintucci S.}{
  address={Harvard-Smithsonian Center for Astrophysics,
60 Garden Street, Cambridge, MA 02138, USA}
}

\author{David L.}{
  address={Harvard-Smithsonian Center for Astrophysics,
60 Garden Street, Cambridge, MA 02138, USA}
}

\author{Vrtilek J.}{
  address={Harvard-Smithsonian Center for Astrophysics,
60 Garden Street, Cambridge, MA 02138, USA}
}

\author{Raychaudhury S.}{
  address={University of Birmingham,
Edgbaston, Birmingham B15 2TT, UK}
}

\author{Nulsen P.}{
  address={Harvard-Smithsonian Center for Astrophysics,
60 Garden Street, Cambridge, MA 02138, USA}
}

\begin{abstract}
  As a part of an ongoing study of a sample of galaxy groups showing
  evidence for AGN/hot gas interaction, we report on the preliminary
  results of an analysis of new {\it XMM} and {\it GMRT} data of the
  X-ray bright compact group HCG 62. This is one of the few groups
  known to possess very clear, small X-ray cavities in the inner
  region as shown by the existing {\it Chandra} image.  At higher
  frequencies ($\ge$1.4 GHz) the cavities show minimal if any radio
  emission, but the radio appears clearly at lower frequencies
  ($\le$610 MHz). We compare and discuss the morphology and spectral
  properties of the gas and of the radio source. We find that the
  cavities are close to pressure balance, and that the jets have a
  "light" hadronic content. By extracting X-ray surface brightness and
  temperature profiles, we also identify a shock front located around
  35 kpc to the south-west of the group center.
\end{abstract}

\maketitle


\section{Introduction}

The work presented here is part of an ongoing project aimed at
combining X-ray and low-frequency radio observations of galaxy groups.
In particular, we have selected a compilation of 18 galaxy groups
based on the presence of signs of interaction between the hot gas and
the central AGN. For these groups, which all have already good quality
X-ray data in the archives of {\it Chandra} and/or {\it XMM}, we have
obtained new radio data at the Giant Metrewave Radio Telescope ({\it
  GMRT}) at 610 MHz and below (Giacintucci et al. 2008, and these
proceedings; Raychaudhury et al. 2009) in order to investigate the AGN
feedback process.

HCG 62 (z=0.0137), the X-ray brightest of the Hickson compact groups,
was the first galaxy group with a clear detection of inner cavities
(Vrtilek et al. 2002).  The existing 1.4 GHz VLA observations mainly
show the emission from the compact radio source and indicate only some
hints of extended radio emission toward the southern cavity. Due to
the poor radio images available, the HCG 62 cavity system was
classified as ``radio ghost'' in the sample of B\^irzan et al. (2004).

We present here new low frequency radio observations of HCG 62 that
allow us to study the X-ray/radio interaction in detail.  With $H_0 =
70 \mbox{ km s}^{-1} \mbox{ Mpc}^{-1}$, and
$\Omega_M=1-\Omega_{\Lambda}=0.3$, its luminosity distance is 59 Mpc
and 1 arcsec corresponds to 0.28 kpc.

\section{X-ray / Radio Interaction}

\subsection{X-ray and radio observations}

For our analysis, we use the X-ray data available in the archives of
{\it Chandra} and {\it XMM}.  In the $\sim$50 ks {\it Chandra} image
the two cavities are clearly visible, but we also notice a
discontinuity in the surface brightness distribution toward the
south-west (SW) direction that we interpret as a shock front (see
below). The combination of this exposure with a more recent
observation of $\sim$100 ks obtained with {\it XMM} allows us to
determine with unprecedented accuracy the X-ray properties of the hot
gas.

We observed HCG 62 with the {\it GMRT} in February 2008 for an
effective time of 2h at 610 MHz and 235 MHz (project 13SGa01).  At
both frequencies, we detect clearly extended emission emanating from
the core in the form of two radio lobes pointing toward the northern
(N) and southern (S) cavities. The overlay of the 235 MHz radio
contours on the smoothed {\it Chandra} image is shown in Fig. 1
(left), whereas the overlay of the 610 MHz radio contours on the XMM
temperature map (see below) is shown in Fig. 1 (right). It is evident
that far more extensive structures become visible at lower
frequencies: the radio emission at 235 MHz is more extended, with two
faint regions located just outside the cavities and apparently bent
toward the east direction.

This radio source is classified as a weak FR-I ($P_{\rm 235 \, MHz}
\sim 2 \times 10^{22}$ W Hz${^{-1}}$), and its spectral index\footnote{
$S_{\nu} \propto \nu^{- \alpha}$ } 
$\alpha^{\rm 610 \, MHz}_{\rm 235 \, MHz} \sim 1.4$ 
is steep compared to that of typical radio galaxies.

\subsection{Energy budget and pressure}

Cavities (and the associated shock fronts, see below) act
as calorimeters for the total energy output of the AGN.  
The total cavity power of HCG 62 is estimated from the X-ray data to
be $\sim 4 \times 10^{42}$ erg s$^{-1}$ (Rafferty et al. 2006).  This
is about twice the luminosity of the ICM inside the cooling region, so
the AGN outburst should be currently supplying enough energy to
balance the cooling in this system.  In fact, since the 235 MHz radio
emission extends beyond the cavities we may argue that there is more
power in the jet and lobes than one can infer from the X-rays.  The
cavity volumes may be larger as the radio emission is less sensitive
to projection effects than depressions in the X-ray image (as also
pointed out by B\^irzan et al. 2008).

Thanks to the new good-quality radio data, we can compare the AGN
mechanical power with the radio luminosity of the source in order to
directly estimate its radiative efficiency.
By assuming a spectral index $\alpha = 1.4$, we measure the total
radio luminosity over the frequency range of 10 MHz - 10 GHz to be
$\sim 4 \times 10^{38}$ erg s$^{-1}$ , which is about four orders of
magnitude less than the total power of the cavities.  Therefore the
radio source in HCG 62 has a very low synchrotron radiative efficiency
($\sim 10^{-4}$).

Since the radio source is filling the cavities, we can also compare
directly the radio pressure internal to the lobes with the X-ray
pressure of the surrounding thermal gas.  The pressure of the thermal
gas is measured from the density and temperature derived from the
X-ray data.  The radio pressure can be estimated under the assumption
that the relativistic plasma is in equipartition with the magnetic
field.  In particular, we assume the so-called ``revised''
equipartition conditions (Brunetti et al. 1997) that include also the
contribution of the low-energy electrons (the Lorentz factor $\gamma$
is as low as 100).  We estimate that the ratio of X-ray pressure to
radio pressure is a factor of a few, and in particular the S cavity is
very close to pressure balance. On the other hand, we can also
determine the ratio $k$ of the energy in protons to that in electrons
that is required to achieve pressure balance (e.g,. Dunn et al. 2005,
B\^{\i}rzan et al. 2008), finding that we have to allow for hadronic
jets with a relatively low value of $k$ ($\sim$10-30).

\begin{figure*}
  \includegraphics[height=.3\textheight]{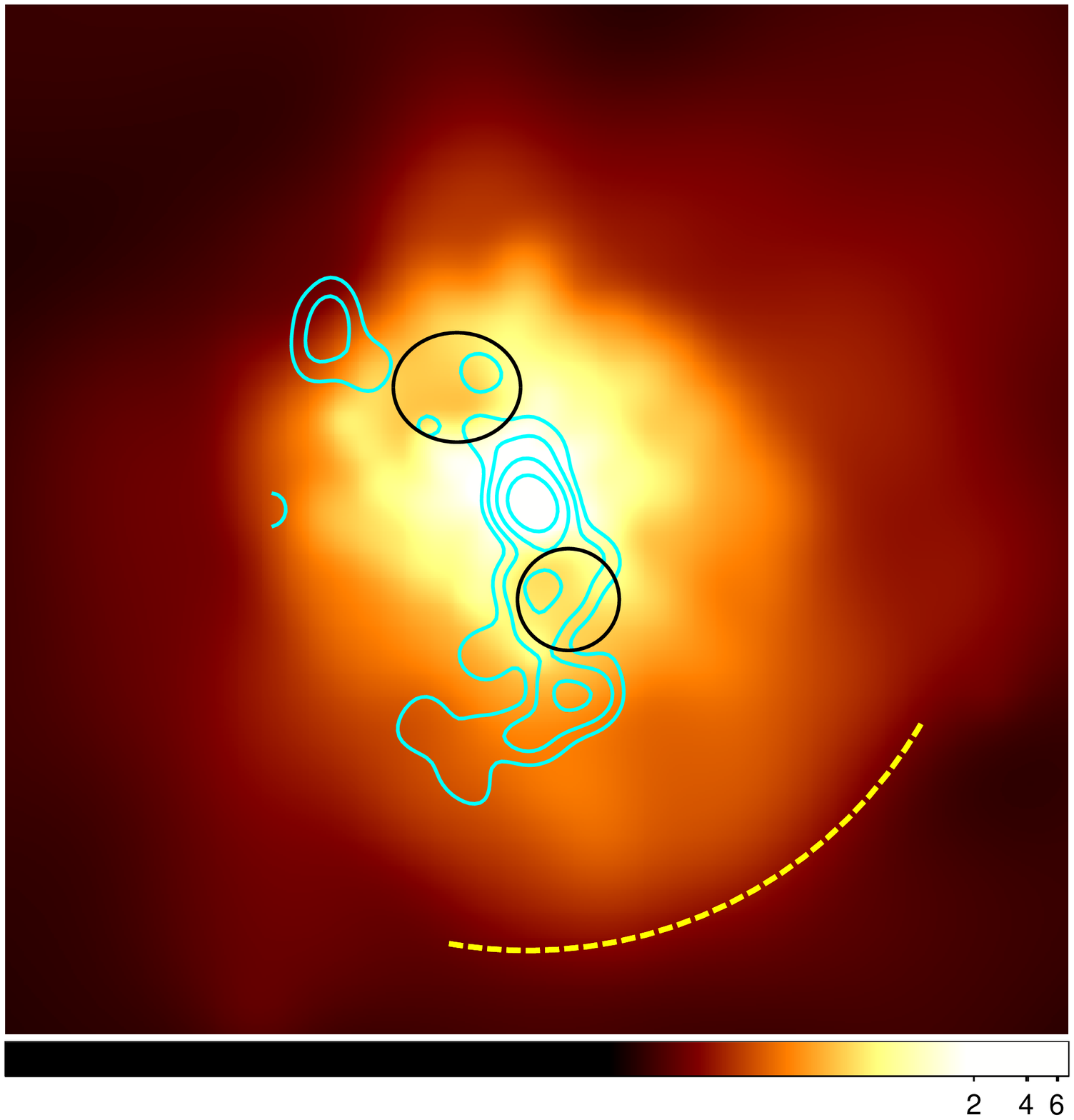}
  \includegraphics[height=.3\textheight]{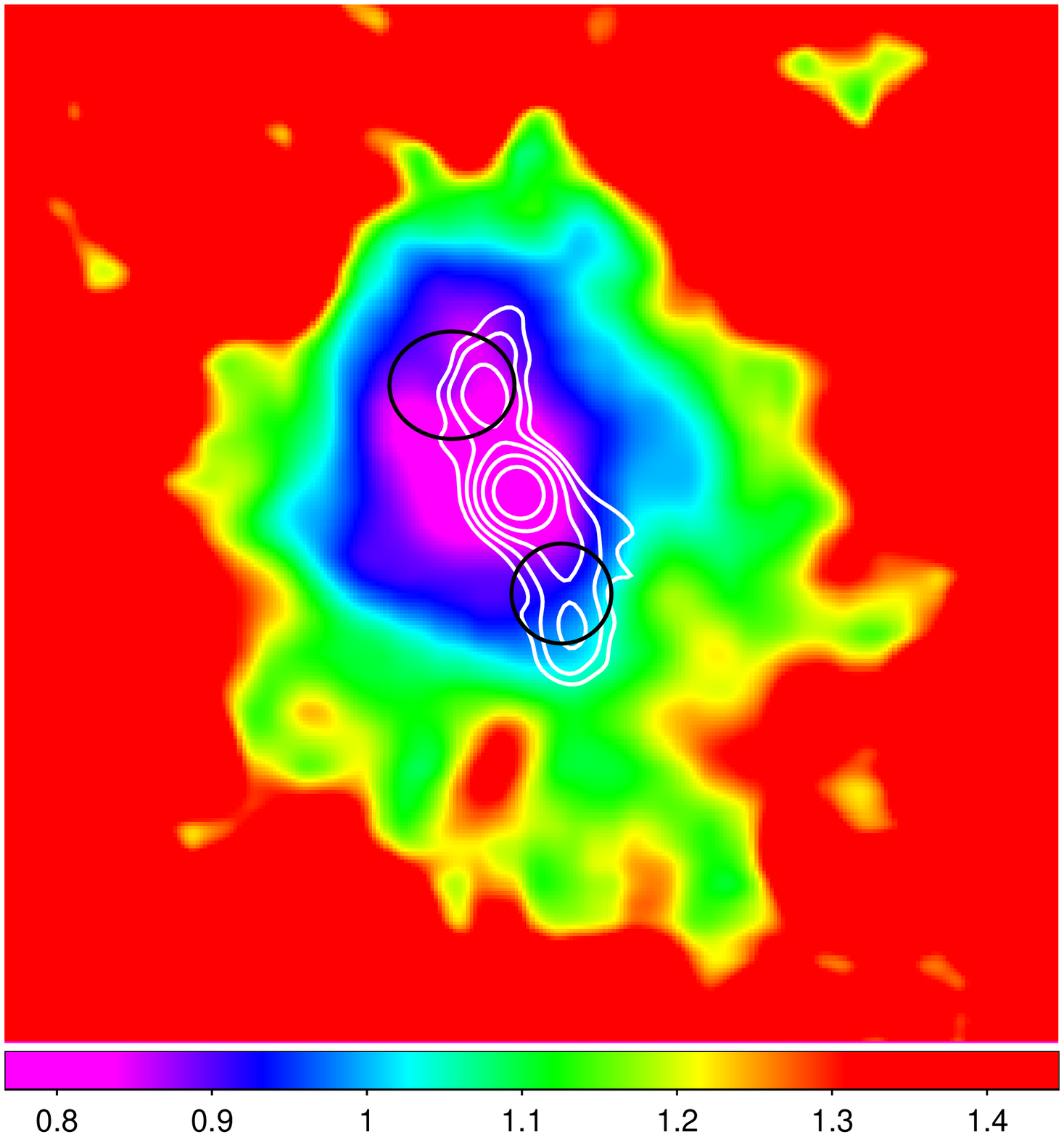} 
  \caption{ \textit{Left:} {\it GMRT} 235 MHz contours overlaid on the
    smoothed 0.5-2.0 keV {\it Chandra} image.  The beam size is $14''$ by
    $12''$ and the lowest contour is shown at $3 \sigma = 0.7$
    mJy/beam. The two elliptical regions and the dashed arc indicate
    the X-ray cavities and the shock front, respectively.
    \textit{Right:} {\it GMRT} 610 MHz contours overlaid on the {\it XMM}
    temperature map.  The beam size is $14''$ by $14''$ and the lowest
    contour is shown at $3 \sigma = 0.2$ mJy/beam. Superposed are the
    cavity regions.  Both boxes are $4'$ by $4'$.}
\end{figure*}

\begin{figure*}
  \includegraphics[height=.3\textheight]{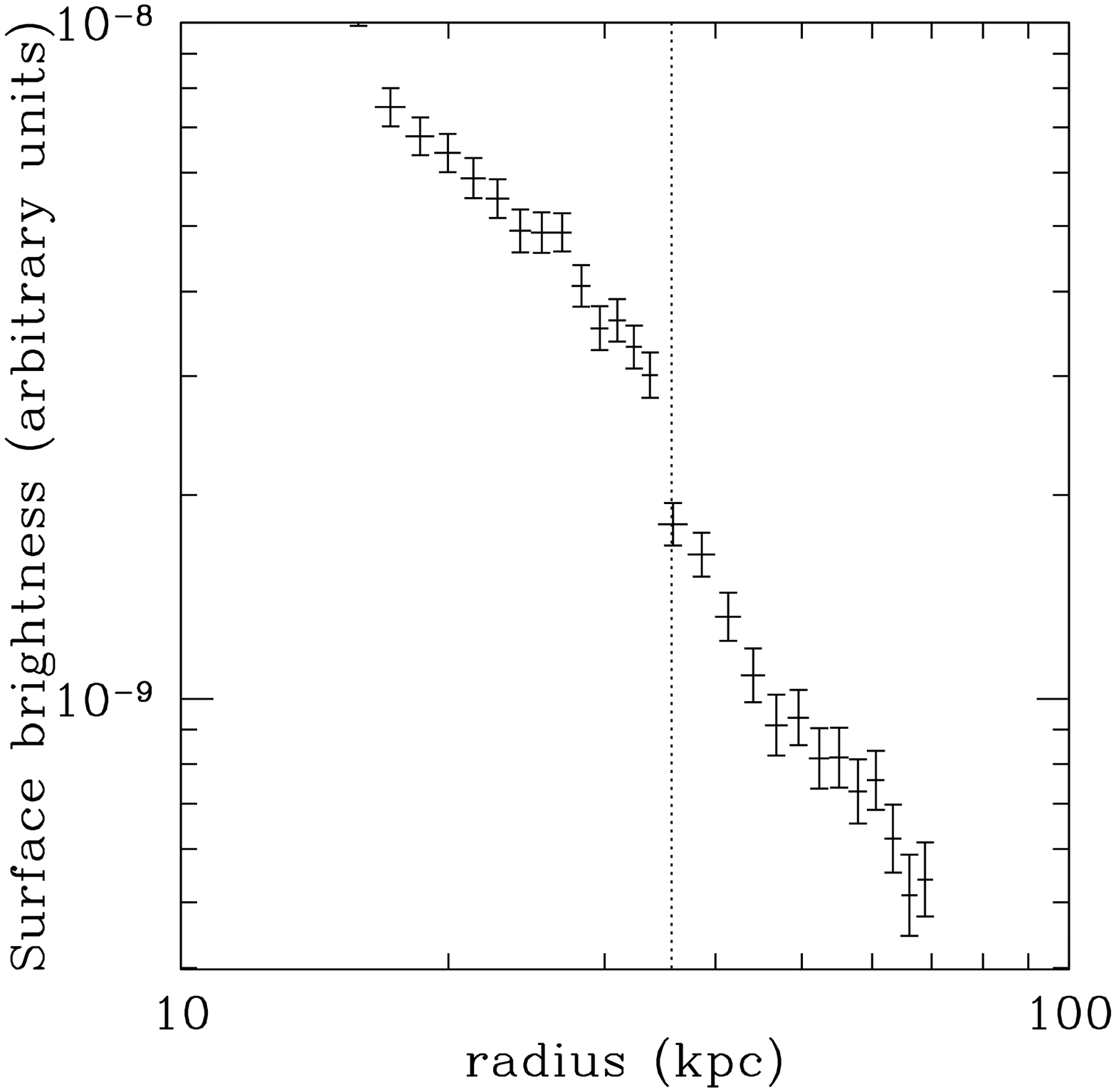}
  \includegraphics[height=.3\textheight]{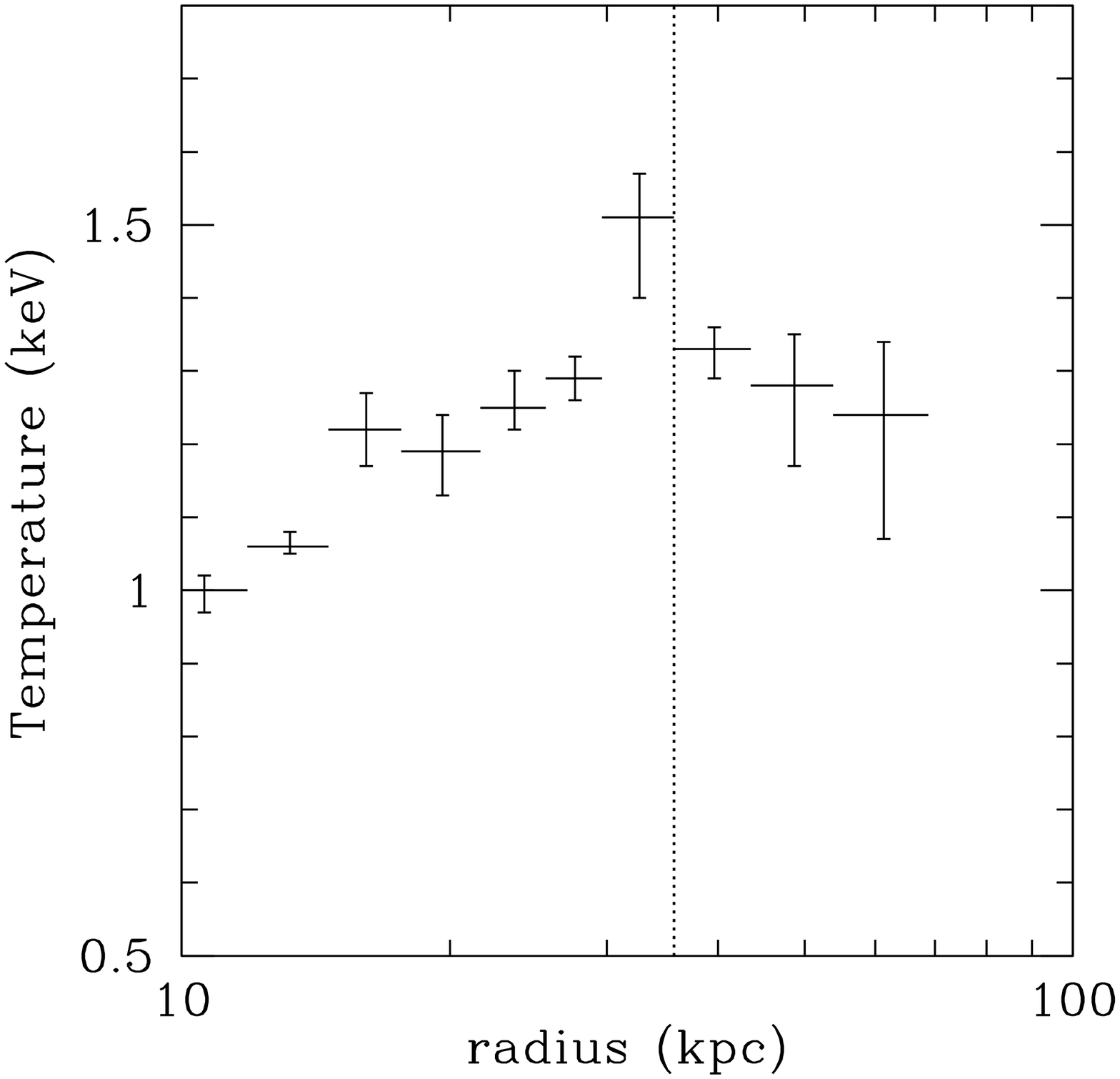}
  \caption{
{\it Left:}
Background-subtracted, exposure-corrected {\it Chandra} surface brightness 
profile extracted along the SW direction (sector between P.A. 
$260^{\circ}$ and $330^{\circ}$). The dotted line indicates the position
of the shock front.
{\it Right:}
{\it Chandra} temperature profile measured in the same sector as in the 
left panel.
}
\end{figure*}

\subsection{Shock front and temperature distribution}

The surface brightness profile extracted along the SW sector (Fig. 2,
left) shows a clear jump at $\sim 35$ kpc from the center.  We derive
the temperature profile along the same sector (Fig. 2, right) and find
that the region immediately interior of the front is significantly
hotter than the undisturbed region just outside of it, with a
temperature jump across the front of $\sim$ 14\%.  Our data are
consistent with the presence of a shock having a Mach number = 1.45,
an energy $\sim 6 \times 10^{57}$ erg, and a power which is about
twice the cavity power.  
The position of the front, just outside the S radio lobe, makes
plausible the interpretation of the shock as being directly driven by
the lobe expansion triggered by the AGN outburst. If so, the cavity
power alone provides a lower limit to the true total mechanical power
of the AGN, as the shock will add to this.

The shock might explain the lack of very cool ($<$0.9 keV) material on
the SW side of the core.  In Fig. 1 (right) is shown the temperature
map obtained with {\it XMM} by using four X-ray colors and estimating
the expected count rate with XSPEC for a thermal {\ttfamily mekal}
model.  Besides the cool core, we notice the presence of a
low-temperature region along the N cavity limbs.  The lack of a
comparable feature at the position of the S cavity edge might imply
that the cool blob extending N of the core once had a symmetric
counterpart on the S side, but this was then heated by the passage of
the shock.

\section{Summary}

\begin{enumerate}

\item Our analysis demonstrates the power of a combined
  X-ray/radio approach to the feedback problem, and particularly the
  benefits of extending radio studies of AGNs to low frequencies where
  less energetic, older electron populations are visible.

\item As opposed to the early classification of the HCG 62 cavity
  system as ``radio ghost'', low-frequency radio emission is detected
  in the cavities by the new {\it GMRT} data.

\item We identify a Mach $\sim$1.5 shock front located $\sim$35 kpc
  to the SW, not reported previously in literature.  The total energy
  in cavities and shock is $\sim 8 \times 10^{57}$ erg.

\end{enumerate}


\begin{thebibliography}{9}

\bibitem[]{} 
B\^{\i}rzan, L. et al.
2004, ApJ, 607, 800

\bibitem[]{} 
B\^{\i}rzan, L. et al.
2008, ApJ, 686, 859

\bibitem[]{}
Brunetti, G. et al. 
1997, A\&A, 325, 898

\bibitem[]{}
Dunn, R. J. H. et al.
2005, MNRAS, 364, 1353

\bibitem[]{}
Giacintucci, S. et al.
2008, ApJ, 682, 186

\bibitem[]{}
Rafferty, D. A. et al.
2006, ApJ, 652, 216

\bibitem[]{}
Raychaudhury, S. et al. 
2009, in The Low-Frequency Radio Universe, arXiv:0907.0895

\bibitem[]{}
Vrtilek, J. M. et al.
2002, American Phys. Soc. Meeting, B17.107

\end{thebibliography}
\end{document}